\documentclass{aa}
\usepackage{graphicx}
\usepackage{natbib}
\bibpunct{(}{)}{;}{a}{}{,}    
\newcommand{\beq}{\begin{equation}}
\newcommand{\eeq}{\end{equation}}
\newcommand{\bea}{\begin{eqnarray}}
\newcommand{\eea}{\end{eqnarray}}

\newcommand{\subscr}[1]{_\mathrm{#1}}

\newcommand{\url}[1]{{\tt #1}}
\newcommand{\pomega}{{\varpi}}
\newcommand{\doverd}[2]{\frac{\partial #1}{\partial #2}}
\newcommand{\doverdt}[1]{\frac{\partial #1}{\partial t}}
\def\gapp{\lower 3pt\hbox{${\buildrel > \over \sim}$}\ }
\def\lapp{\lower 3pt\hbox{${\buildrel < \over \sim}$}\ }

\begin{document}
\title{Evolution of Planetary Systems in Resonance}
\author{
Wilhelm Kley\inst{1}, 
Jochen Peitz\inst{1},
\and
Geoffrey Bryden\inst{2}}
\offprints{W. Kley,\\ \email{kley@tat.physik.uni-tuebingen.de}}
\institute{
     Institut f\"ur Astronomie \& Astrophysik, 
     Abt. Computational Physics,
     Universit\"at T\"ubingen,
     Auf der Morgenstelle 10, D-72076 T\"ubingen, Germany
\and
Jet Propulsion Lab, MS 169-506, 4800 Oak Grove Dr, Pasadena, CA 91109, USA}
\date{Revised:  July, 2003}
\abstract{%
We study the time evolution of two protoplanets
still embedded in a protoplanetary disk.
The results of two different numerical approaches are presented and compared.
In the first approach, the motion of the disk material is computed with
viscous hydrodynamical simulations, and the planetary motion is
determined by N-body calculations including exactly the gravitational
forces exerted by the disk material.
In the second approach, only the N-body integration is performed
but with additional dissipative forces included such as to mimic
the effect of the disk torques acting on the disk.
This type of modeling is much faster than the full hydrodynamical
simulations, and gives comparative results provided that parameters
are adjusted properly.  

Resonant capture of the planets is seen in both approaches, where
the order of the resonance depends on the properties of the disk
and the planets.
Resonant capture leads to a rise in the eccentricity and to an
alignment of the spatial orientation of orbits.
The numerical results are compared with the
observed planetary systems in mean motion resonance
(Gl~867, HD~82943, and 55~Cnc).
We find that the forcing together of two planets by their parent disk
produces resonant configurations similar to those observed,
but that eccentricity damping greater than that obtained
in our hydrodynamic simulations is required to match the
GJ~876 observations.
}
\maketitle
\markboth
{Kley, Peitz, and Bryden: Planetary Systems in Resonance}
{Kley, Peitz, and Bryden: Planetary Systems in Resonance}
\section{Introduction}
\label{sec:introduction}
Since their first discovery in 1995, the number
of detected extrasolar planets
orbiting solar-type stars has risen during recent years to more than 100 
(for an up-to-date list see
e.g. {\tt {http://www.obspm.fr/encycl/encycl.html}} by J.~Schneider).
Among these, there are currently 11 systems with two or more planets;
a summary of their properties has been given recently by
\citet{2002marcy-systems}.
With further observations to come, the fraction of systems with
multiple planets will almost certainly
increase, as many of the systems exhibit long-term trends in their
radial velocity,
suggesting an additional outer planet.
Among the known multiple-planet extrasolar systems there are now three
confirmed cases, namely Gl~876 \citep{2001ApJ...556..296M},
HD~82943 (the {\it Coralie Planet Search Programme},
ESO Press Release 07/01), and 55~Cnc 
\citep{2002ApJ...581.1375M}
where the planets orbit their central star in a low-order
{\it mean motion resonance}
such that the orbital periods have nearly exactly the ratio 2:1 or 3:1.
The parameters of these planetary systems
are displayed in Table~\ref{tab:system} below.
The possibility of a 2:1 resonance in HD~160691 has also been discussed recently
by \citet{2003astro-ph..0301528}, although the orbital periods are too long
to definitely confirm this.
Overall, these numbers imply that at least one-fourth of multiple-planetary systems
contain planets in resonance,
a fraction which is even higher if secular resonances,
such as those observed in the $\upsilon$ And system,
\citep{1999ApJ...526..916B} are also considered.

The formation of resonant planetary systems can be understood
by considering the joint evolution of protoplanets together with the
protoplanetary disk from which they formed.
Using local linear analysis, it has been shown
that the gravitational interaction of a single protoplanet with
its disk leads to torques resulting in a change of the semi-major
axis (migration) of the planet
\citep{1980ApJ...241..425G, 1986ApJ...309..846L,
 1997Icar..126..261W, 2002ApJ...565.1257T}.
Additionally, as a result of angular momentum
transfer between the viscous disk and the planet,
planetary masses of around one Jupiter mass can open gaps in the
surrounding disk \citep{1980MNRAS.191...37L, 1993prpl.conf..749L}.
Fully non-linear hydrodynamical calculations 
for Jupiter-sized planets 
\citep{1999MNRAS.303..696K, 1999ApJ...514..344B, 
1999ApJ...526.1001L, 2000MNRAS.318...18N, 2002A&A...385..647D}
confirmed this expectation and clearly showed
that disk-planet interaction leads to:
{\it i)} excitation of spiral shock waves in the disk, whose tightness
depends on the sound-speed in the disk,
{\it ii)} formation of an annular gap,
whose width is determined by the balance
between gap-opening tidal torques and gap-closing viscous plus pressure
forces,
{\it iii)} inward migration on a time scale of $10^5$ yrs for typical
disk parameters, in particular disk masses corresponding to that of the
minimum mass solar nebula,
{\it iv)} possible mass growth after gap formation up to about
10 $M_{Jup}$ when finally the gravitational torques overwhelm the
diffusive tendencies of the gas, and
{\it v)} a prograde rotation of the planet.
New three-dimensional computations with high resolution resolve the
flow structure in the vicinity of the planet,
and allow for more accurate estimates of the mass accretion and
migration rates \citep{2003ApJ...586..540D, 2003astro-ph..0301154}.

These hydrodynamic simulations with single planets
have been extended to models which contain
multiple planets.
It has been shown
\citep{2000MNRAS.313L..47K,2000ApJ...540.1091B,
2001A&A...374.1092S,2002MNRAS.333L..26N}
that during the early evolution,
when the planets are still embedded in the disk,
different migration speeds may lead to an approach of neighboring
planets and eventually to resonant capture.
More specifically, the evolution of planetary systems into a 2:1
resonant configuration was seen in the calculations of 
\citet{2000MNRAS.313L..47K}
prior to the discovery of any such systems.

In addition to hydrodynamic disk-planet simulations, many authors have analyzed
the evolution of multiple-planet systems with N-body methods.
Each of the known resonant systems have been considered in detail.
\citet{2002ApJ...572.1041J} and \citet{2002ApJ...567..596L}
have modeled the evolution of 2:1 resonant system GJ~876, while
the 3:1 system 55~Cnc has been analyzed by
\citet{2003ApJ...585L.139J} and \citet{2002astro-ph..0209176},
and the 2:1 system HD 82943 by
\citet{2001ApJ...563L..81G} and \citet{2002astro-ph..0301353}.
Based on orbit integrations, 
these papers confirm that the planets in these systems are 
in resonance with each other.
The dynamics and stability of resonant planetary
systems in general has been recently studied by
\citet{2002astro-ph..0210577}.

Here we present new numerical calculations treating the
evolution of two planets still embedded in a protoplanetary disk.
We use both hydrodynamical simulations and simplified N-body
integrations to follow the evolution of the system. In the first approach,
the disk is evolved by solving the full time-dependent
Navier-Stokes equations simultaneously
with the evolution of the planets. Here, the motion of the planets
is determined by the gravitational action of both planets,
the star, and the disk.
In the latter approach, we take a simplified
approximation and perform 3-body (star plus two planets) calculations
augmented by additional (damping) forces which approximately
account for the gravitational influence of the disk \citep[e.g.][]{2002ApJ...567..596L}.
Using both approaches, allows a direct comparison of the 
alternative methods,
and does enable us to determine the damping parameters
required for the simpler (and much faster) second type of approach.

\section{The Observations}
\label{sec:obs}
The basic orbital parameters of the three known systems in mean motion resonance
are presented in Table~\ref{tab:system}. 
The orbital parameters for Gl~876 are taken from the
dynamical fit of \citet{2001ApJ...551L.109L}, and for
HD~82943 from \citet{2001ApJ...563L..81G}.
Due to the uncertainty in the inclinations of the systems, 
$M\sin i$, rather than the exact mass of each planet, is listed.
By including the mutual perturbations of the planets into their fit 
of Gl~876, \citet{2001ApJ...551L.109L}, however, 
are able to constraint that system's inclination to $\sim 30^o - 50^o$.

Two of the systems,
Gl~876 and HD~82943, are in a nearly exact 2:1 resonance.
We note that in both cases the outer planet is more massive,
in one case by a factor of about two (HD~82943) and in the
other by more than three (Gl~876).
The eccentricity of the inner (less massive) planet is larger than
that of the outer one in both systems. For the system Gl~876 the alignment of the orbits
is such that the two periastrae are pointing in nearly the same
direction. For the system HD~82943 these data have not been
clearly identified, due to the much longer orbital periods,
but they do not seem to be very different from each other.
The third system, 55~Cnc, is actually a triple system. Here the inner
two planets orbit the star very closely and are in a 3:1
resonance, while the third, most massive planet orbits at a
distance of several AU.
\begin{table}
\caption{
The orbital parameters of the three systems known to contain a mean
motion resonance. $P$ denotes the orbital period,  $M\sin i$ the
mass of the planets, $a$ the semi-major axis, $e$ the eccentricity,
$\pomega$ the angle of periastron, and $M_*$ the mass of the central
star. It should be noted that the orbital elements for shorter period
planets undergo secular time variations. Thus in principle one should
always state the epoch corresponding to these osculating elements
\citep[see e.g.][]{2001ApJ...551L.109L}.
}
\label{tab:system}
\begin{tabular}{c|l|l|l|l|l|l} 
\hline
   Name  &   P  & $M\sin i$  &  a        &   $e$  &  $\pomega$ & $M_*$  \\
       &   [d]  & [$M_{Jup}$]  &  [AU]   &        &   [deg] &  [$M_\odot$]   \\
 \hline
   Gl~876& (2:1)  &   &   &  &   &  0.32  \\
   c   &  30.1 &   0.56         &  0.13   &   0.24  &  159  &    \\
   b   &   61.02  & 1.89           &  0.21  &   0.04  &  163  &     \\
 \hline
   HD~82943 & (2:1)  &   &   &  &   & 1.05  \\
   b   &  221.6  &  0.88          &  0.73   &   0.54 &   138 &    \\
   c   &  444.6  & 1.63          &  1.16   &   0.41 &    96 &    \\
 \hline
   55~Cnc & (3:1) &   &   &  &   &  0.95   \\
   b   &  14.65 &  0.84          &  0.11   &   0.02 &   99  & \\
   c   &  44.26 &   0.21         &  0.24   &   0.34 &    61 &   \\
    d   & 5360   & 4.05          &  5.9    &   0.16 &   201 &    \\
 \hline
\end{tabular}
\end{table}
\section{The Models}
\label{sec:model}
Our goal is to investigate the evolution of protoplanets
still embedded in their disk.
As outlined in the introduction, we employ two
different methods which complement each other.  First, a fully
time-dependent hydrodynamical model for the joint evolution
of the planets {\it and} disk is presented.
Because the evolutionary time scale may cover several thousands of orbits,
the fully hydrodynamical computations (of disk and planets)
can require millions of time-steps, which
translates into an effective computational time of up to several
weeks.

Since our main interest is the orbital
evolution of the planets and not so much the hydrodynamics of the
disk, we also perform 3-body orbit integrations which do not
explicitly follow the disk's evolution. 
Through a direct comparison with the hydrodynamical models,
it is then possible to infer the effective damping forces
to include within this faster calculation.
\subsection{Hydrodynamical Model}
\label{subsec:model-fullhydro}
The first set of coupled hydrodynamical-N-body
models presented in this paper are calculated in the
same manner as the models described previously in
\citet{1998A&A...338L..37K, 1999MNRAS.303..696K}
for single planets and in \citet{2000MNRAS.313L..47K}
for multiple planets. The reader is referred to those papers
for details on the computational aspects of the simulations.
Other similar models, following explicitly the motion of single and
multiple planets in disks, have been presented by 
\citet{2000MNRAS.318...18N}, \citet{2000ApJ...540.1091B}, and \citet{2001A&A...374.1092S}.
\subsubsection{Equations}
For reference, we summarize the basic equations for this
problem.
We use two-dimensional cylindrical coordinates ($r, \varphi$),
where $r$ is the radial coordinate and $\varphi$ is the azimuthal angle.
Thus, we consider an infinitesimally thin disk located at $z=0$,
with a velocity field ${\bf u} = (u_r, u_{\varphi})$.
The origin of the coordinate system is at the position of
the star.
In the following we will use the symbol $v=u_r$ for the radial velocity
and $\omega = u_{\varphi}/r$ for the angular velocity of the flow.
As there is no preferred rotational frame, we work in a non-rotating
reference system.
Then the equations of motion are
\begin {equation}
 \doverdt{\Sigma} + \nabla \cdot (\Sigma {\bf u} )
                =  0,  \label{eq:Sigma}
\end{equation}
\begin {equation}
 \doverdt{(\Sigma v)} + \nabla \cdot (\Sigma v {\bf u} )
  = \Sigma \, r \omega^2
        - \doverd{p}{r} - \Sigma \doverd{\Phi}{r} + f\subscr{r}
      \label{eq:u_r}
\end{equation}
\begin {equation}
 \doverd{(\Sigma r^2 \omega)}{t}
   + \nabla \cdot (\Sigma r^2 \omega {\bf u} )
         =
        - \doverd{p}{\varphi} - \Sigma \doverd{\Phi}{\varphi}
   + f\subscr{\varphi}
      \label{eq:u_phi}
\end{equation}
Here $\Sigma$ denotes the surface density, $p$ the
two-dimensional pressure, and $f_r, f_\varphi$ denote the
components of the viscous forces, given explicitly in
\citet{1999MNRAS.303..696K}. The
gravitational potential $\Phi$ generated by the protostar
with mass $M_*$ and the two planets having mass $m_1$ and $m_2$
is given by
\begin{equation}  \label{eq:Phi}
    \Phi = - \frac{G M_*}{| {\bf r} - {\bf r}_* |}
       -  \frac{G m_1}{ \left[ ( {\bf r} - {\bf r}_1 )^2
             + s_{1}^2 \right]^{1/2} }
       -  \frac{G m_2}{ \left[ ( {\bf r} - {\bf r}_2 )^2
             + s_{2}^2 \right]^{1/2} }
\end{equation}
where $G$ is the gravitational constant and ${\bf r}_*$,
${\bf r}_1$, and ${\bf r}_2$ are the position vectors for the
star and the inner/outer planet, respectively.
The quantities $s_1$ and $s_2$ are smoothing lengths
that are set to 1/5 of each planet's Roche radius.
This softening of the potential removes any local fluctuations that
might result as the planets move through the computational grid.

The motion of the star and the planets is determined firstly by their
mutual gravitational interaction and secondly by the gravitational
forces exerted upon them by the disk.
The acceleration of the star ${\bf a}_*$ is given for example by
\begin{eqnarray}  \label{a_*}
    {\bf a}_* =
        G m_1 \frac{ {\bf r}_1 - {\bf r}_*}{ | {\bf r}_1 - {\bf r}_* |^3}
     & + &  G m_2 \frac{ {\bf r}_2 - {\bf r}_*}{ | {\bf r}_2 - {\bf r}_* |^3}
         \nonumber \\
     & + & G \int_{Disk} \Sigma (r,\varphi) \, \frac{ {\bf r} - {\bf r}_*}
              { | {\bf r} - {\bf r}_* |^3} \,  dA
\end{eqnarray}
where the integration covers the entire disk surface.
The accelerations of the planets are found similarly.
We work here in an accelerated coordinate frame where the origin
is located in the center of the (moving) star.
Thus, in addition to force due to the gravitational potential
(Eq.~\ref{eq:Phi}), the disk and planets also feel an acceleration $-{\bf a}_*$.
\subsubsection{Initial and boundary conditions}
The initial hydrodynamic structure of the disk, which extends radially
from $r_{min}$ to $r_{max}$,
is axisymmetric with respect to the location of the star, and
the surface density scales as  $\Sigma(r) = \Sigma_0 \, r^{-1/2}$, with
superimposed initial gaps \citep{2000MNRAS.313L..47K}. 
The initial velocity is pure Keplerian rotation ($v_r=0,
v_\varphi = G M_*/r^{1/2}$). We assume a fixed temperature
law with $T(r) \propto r^{-1}$ which follows from the assumed
constant vertical height $H/r$.
The kinematic viscosity $\nu$ is typically parameterized by an 
$\alpha$-description $\nu = \alpha c_s H$,
with the sound speed $c_s = H v_\varphi / r$, only one
model (X) has a constant kinematic viscosity.

The radial outer boundary is closed, i.e. $v_r(r_{max}) = 0$.
At the inner radial boundary outflow boundary conditions are
applied; matter may flow out, but none is allowed to enter. This
procedure mimics the accretion process onto the star. The density gradient
is set to zero at $r_{min}$ and $r_{max}$, while the
angular velocity there is fixed to be Keplerian.
In the azimuthal direction, periodic boundary conditions for all
variables are imposed.
\subsubsection{Model parameters}
We present several models that are listed in 
Table~\ref{tab:modpar},
for the complete model parameters see below.
In all cases, the planets are allowed to migrate
(change their semi-major axes)
through the disk in accordance with the gravitational
torques exerted on them. 
During the evolution, material is removed from the centers
of the planets' Roche-lobes and is
assumed to have been accreted onto each planet, for the
detailed procedure see \citet{1999MNRAS.303..696K}.
However, in spite of this assumed accretion process,
in most models this mass is not added to the dynamical mass
of the planets.
Hence, the disk mass is slowly depleted while the
planets' masses are held fixed at the initial
values.
This assumption of constant planet mass throughout the
computation is well justified, as the
migration rate depends only weakly on the mass of the planet
\citep{2000MNRAS.318...18N}.
Only in the last model X are the masses of the planets allowed
to grow during the computation, which may test this assumption. 
\begin{table}
\caption{
Planetary and disk parameters of the hydrodynamic models.
The masses of the planets are
given in Jupiter masses ($M_{Jup} = 10^{-3} M_{\odot}$).
Index 1 refers to the inner and index 2 to the outer planet.
For the last model X, we allow for variable planet masses which
both start with 1 $M_{Jup}$ and then grow during the computation.
For the viscosity, the corresponding value of $\alpha$ is given except
for the last case (model X) in which we use a constant kinematic viscosity
(corresponding to $\alpha=0.04$ at a radial location of $5.2$ AU).
The relative disk scale height $H/r$ is given in the last column.
For the disk masses and computational domains,
see section \ref{subsec:model-fullhydro}.
}
\label{tab:modpar}
\begin{tabular}{ccccl}
 \hline
Model &   $m_1$  &  $m_2$  &    Viscosity  & $H/r$   \\ 
     & [$M_{Jup}$] &  [$M_{Jup}$]  &  $\alpha$     \\
 \hline
  A    &    5    &    3    &  $10^{-2}$   &   0.10   \\
  B    &    5    &    3    &  $2 \cdot 10^{-3}$   &   0.10   \\
 \hline
  C    &    3    &    5    &  $10^{-2}$   &   0.10   \\
 D1, D2 &  3    &    5    &  $3 \cdot 10^{-3}$   &   0.10  \\ 
  E    &    3    &    5    &  $10^{-3}$   &   0.10  \\
  F    &    3    &    5    &  $10^{-2}$   &   0.075 \\
  G    &    3    &    5    &  $10^{-2}$   &   0.05  \\ 
 \hline
  X    &   1 (V)   &    1 (V)   &    $\nu = const.$ &  0.05  \\
 \hline
\end{tabular}
\end{table}

In all the models A-G, the disk extends radially
from  $r_{min} = 1$ to $r_{max}= 30$ AU. The disk mass within this
radial range is about 0.04 $M_\odot$, and the planets are always placed
initially at 4 and 10 AU, respectively. In the first two models (A \& B),
the inner planet is more massive than the outer one, while in models
C through G the inner planet is less massive.  A range of
values for the viscosity and disk thickness $H/r$ are considered,
as listed in Tab.~\ref{tab:modpar}.
The values of $\alpha =0.01$ and $H/r =0.1$
may be on the high side
for protoplanetary disks but allow for a sufficiently rapid
evolution of the system to identify clearly the governing physical
effects. The influence of these parameters is studied by comparison between
different models.
The parameters for models D1 and D2 are identical, but in the second 
case the density has been perturbed randomly by 1\%.

Model X differs in several respects from the other models.
The kinematic viscosity $\nu$ is constant and equivalent to 
$\alpha =0.004$ at 5.2 AU.
Here, both embedded planets each have initial masses of 1 $M_{Jup}$ and
are placed initially at 1 and 2 AU.
This model is a continuation of the one presented previously in \citet{2000MNRAS.313L..47K}.
The radial extent of the disk for this model was 
$r_{min} = 1.3$ to $r_{max}= 20.8$ AU, which contained a disk mass
of $0.01 M_{\odot}$ initially. The initial setup
makes the surface density $\Sigma(r)$ the same for all models.
\subsection{Damped N-body Model}
\label{subsec:model-Nbody}
As pointed out, the full hydrodynamical evolution is computationally very
time consuming since tens of thousands of orbits must be calculated.
Hence, we also perform simpler N-body calculations
of planetary systems with two planets orbiting a single star.
We consider only {\it coplanar} systems, where the planetary orbits
and the equatorial plane of the disk are all aligned.
The orbit of a single planet around a star is an ellipse and
is determined by three orbital elements: the semi-major axis $a$, the
eccentricity $e$, and the direction of
periastron $\pomega$. The actual location of the
planet within the orbit can be obtained
from the elapsed time since last periastron.
In the case of a planetary system with more than one planet,
due to the mutual gravitational perturbations 
the ellipses are no longer fixed in space.
In the case where the masses of the planets are
much smaller that the stellar mass, at each epoch we can fit an
instantaneous ellipse to the orbit of each planet and obtain the
corresponding osculating elements of the orbit.
These are calculated for each planet individually, considering only
one planet at a time.

The gravitational influence of the surrounding disk is modeled
here through prescribed (damping) forces.
We assume that these forces act on the momentary
semi-major axis and eccentricity of the planets
through explicitly specified relations for $\dot{a} (t)$ and $\dot{e} (t)$,
which vary with time.
The changes $\dot{a}$ and $\dot{e}$
caused by the damping effects of the disk can be translated into
additional forces changing directly the position ${\vec{x}}$
and velocity ${\vec{u}}$ of the planets. Our implementation
follows \citet{2002ApJ...567..596L}, where 
explicit expressions for these damping terms are given.
As a test, we recalculated their model for
GJ~876 and confirmed their results.

\begin{figure}[ht]
\begin{center}
\resizebox{0.98\linewidth}{!}{%
\includegraphics{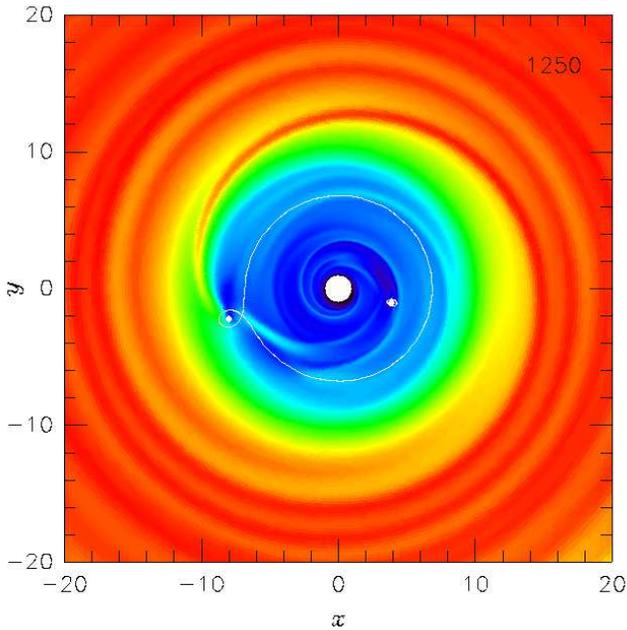}}
\end{center}
  \caption{
  Overview of the density distribution of model C
  after 1500 orbital periods of the inner planet.
  Higher density regions are brighter and lower ones are darker.
  The star lies at the center of the white inner region 
  bounded by $r_{min}=1$ AU. The location of the two planets is
  indicated by the
  white dots, and their Roche-lobes are also drawn. Clearly seen are
  the irregular spiral wakes generated by the planets. 
  Regular intertwined spiral arms are seen only
  outside of the second planet.
    }
   \label{fig:overview}
\end{figure}
Motivated by the basic idea of two planets orbiting inside a disk's cavity
(see Fig.~\ref{fig:overview}), we damp $a$ and $e$ for
the {\it outer} planet only.
We adopt a logarithmic time derivative of $a$ of the form
\beq
   \frac{\dot{a}}{a}  
   = - \frac{1}{\tau(t)} \, \qquad \mbox{with}
    \qquad  \,  \tau(t) = \, \tau_0  \, + \, \beta t,
\label{eq:adot}
\eeq
where $\tau(t)$ denotes the damping time, and $\beta$ a dimensionless 
positive 'stretching' constant.
By making the ansatz Eq.~(\ref{eq:adot}) we tried to make the 
damping as simple as possible with only two parameters to fit. 
In practice we found that 
the damping time $\tau$ could not be chosen as a fixed constant, 
such that we assume a simple linear time dependence.
Equation~(\ref{eq:adot}) can be integrated to yield
\beq
   a(t) = a_0 \cdot 
   \left( 1 \, + \, \beta \, \frac{t}{\tau_0} \right)^{- 1/\beta},
\label{eq:a}
\eeq
where ${a}_0$ denotes the starting value of $a$
at the initial time $t_0=0$. 

The eccentricity damping is set to a fixed multiple of the
semi-major axis damping
\beq
   \frac{\dot{e}}{e}  =  K \, \, \frac{\dot{a}}{a} ,
\label{eq:edot}
\eeq
where $K$ is a constant, typically larger than unity
\citep[see also][]{2002ApJ...567..596L}.

The time scale $\tau_0$, the 'stretching' factor $\beta$,
and $K$ are adjusted in order to match the results
of the full hydrodynamic calculations.
Results of the two methods
are compared in Section \ref{subsec:Nbody} below.

\section{Results}
\label{sec:results}
The basic evolutionary sequence of two planets evolving
simultaneously with a hydrodynamic disk has been calculated and described by
\citet{2000MNRAS.313L..47K}  and \citet{2000ApJ...540.1091B}.
The model X was considered in \citet{2000MNRAS.313L..47K},
where it was found that both planets evolve into a 2:1
resonant configuration. 
Model X and model A are
discussed further in a recent conference paper
\citet{2002astro-ph..0302352}. Here, these models are listed essentially
for completeness, while our main interest in this paper
focuses on models C-G.

To analyze the system dynamics in the presence
of a disk, we monitor the evolution of the orbital elements
$a$, $e$, and $\pomega$ of both planets throughout the simulations.
In the case of a resonance, the orbits
are coupled dynamically. For coplanar systems that are
in a mean-motion commensurability $(p+q)$:$p$, it suffices to
use two resonant angles to describe the evolution.
Here we consider the angular difference of the apsidal lines
\beq
   \Delta \pomega = \pomega_2 - \pomega_1
\eeq
and the combination
\beq
   \Theta_1  = (p+q) \lambda_2  -  p \lambda_1  - q \pomega_1,
\label{eq:theta}
\eeq
where $\lambda_i$ are the mean longitudes of the inner ($i=1$)
and outer ($i=2$) planets.
The two resonant angles $\Delta \pomega$ and $\Theta_1$ have also
been used recently by \citet{2002astro-ph..0210577} to study the possible
stable solutions of resonant planetary systems.
In Eq.~(\ref{eq:theta}),
we have $p=1$, $q=1$ for a 2:1 resonance, while for a 3:1  
resonance $p=1$, $q=2$.

Before we discuss details of resonant planetary evolution we
briefly summarize the main properties of model C,
which serves as our standard reference model.
\subsection{Hydrodynamical Models}
\label{subsec:fullhydro}
\subsubsection{Overview}
At the start of the simulations both planets are placed into
an axisymmetric disk, where the density is initialized with
partially opened gaps superimposed
on an otherwise smooth radial density profile.
Upon starting the evolution the two main effects are:
\begin{enumerate}
\item[a)]
Because of the accretion of gas onto the two planets the
radial region in between them is depleted in mass and finally
cleared. This clearing time depends on the mass of the planets and on the
viscosity and temperature of the disk. The individual gaps of
higher mass planets are deeper, which lengthens the overall clearing
timscale. Higher
viscosity and temperature tend to 'push' material towards the
planets and shorten the clearing time. Additionally, the disturbance
by the two planets creates a strongly time dependent flow with two
mutually interacting spiral arms which also pushes matter towards
the planets. 
The snapshot in Fig.~\ref{fig:overview} shows clearly this effect, which
again reduces the clearing time.
Thus, the high viscosity ($\alpha = 0.01$) and temperature
($H/R=0.1$) of the standard model (C) still allow for gap clearing in spite
of the large mass of the planets. After about 5000 yrs, only 2\% of the
initial material between the planets remains.
For model X, with lower mass planets,
this phase takes only a few hundred orbital periods
\citep{2000MNRAS.313L..47K}.

Concurrently with the central ring depletion,
the region interior to the inner planet
loses material either due to accretion onto the central star
(the inner boundary is open to outflow) or by accretion onto the planet.
As with the intermediate ring, the
timescale for clearing this inner region again depends on the
physical parameter of the system.
Thus, after an initial transient phase we typically expect the
configuration of two planets
orbiting within an inner cavity of the disk,
as seen in Fig.~\ref{fig:overview} 
\citep[see also][]{2000MNRAS.313L..47K}.
\item[b)]
After initialization, the planets quickly (within a few orbital
periods) excite non-axisymmetric disturbances, viz. the spiral waves,
in the disk. In contrast to the single planet case these are
not stationary in time, because there is no preferred rotating
frame.
The gravitational torques exerted on the two planets
by those non-axisymmetric density perturbations induce a migration
process for the planets.
\end{enumerate}
\begin{figure}[ht]
\begin{center}
\resizebox{0.98\linewidth}{!}{%
\includegraphics{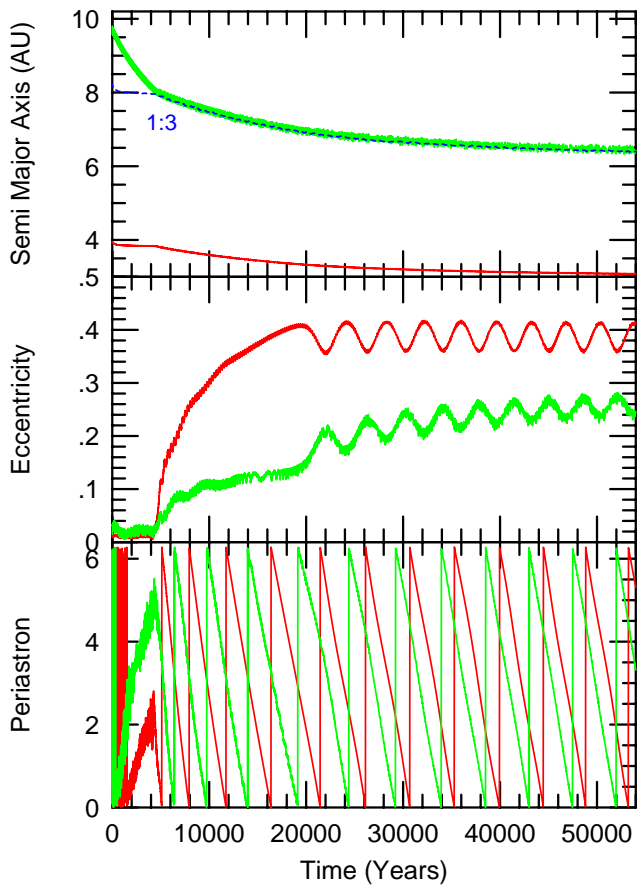}}
\end{center}
  \caption{The semi-major axis ($a$), eccentricity ($e$)
and position angle of the 
orbital periastron ($\pomega$) for the two planets versus time for Model C.
In this example, the planets have fixed masses of
3 and 5 $M_{Jup}$, and are placed initially at 4 and 10 AU,
respectively. The inner planet is denoted by the black line,
the outer by the light gray line. 
The dotted reference line (labeled 3:1), indicates the
location of the 3:1 resonance with respect to the inner
planet.
}
\label{fig:aeo-tw6y}
\end{figure}
\begin{figure}[ht]
\begin{center}
\resizebox{0.98\linewidth}{!}{%
\includegraphics{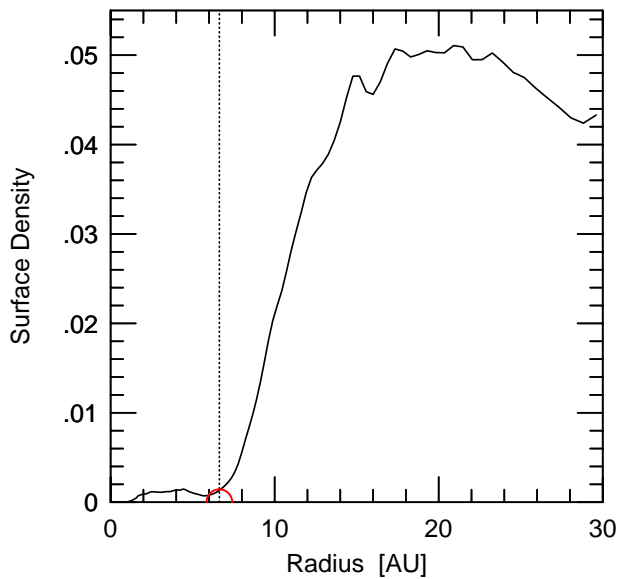}}
\end{center}
\caption{
The azimuthally averaged surface density (in dimensionless units)
for model C at 40,000 yrs. The location of the semi-major axis
of the planet is indicated by the vertical dashed line at $r = 6.63$,
and the size of its Roche-radius by the solid circle. The left and right
arrows indicate the locations of the inner and outer 2:1 Lindblad
resonances, respectively.
\label{fig:sigma}
}
\end{figure}
Now, the planets' relative positions within the cavity
have a distinct influence on their subsequent evolution.
As a consequence of the clearing process, the inner planet is no longer
surrounded by any disk material and thus cannot grow any further
in mass. In addition, it cannot migrate anymore, because
there is no torque-exciting material left in its vicinity.
All the material
of the outer disk is still available, on the other hand, to exert negative 
(Lindblad) torques on the outer planet.
Hence, in the initial phase of the computations we observe an
inwardly migrating outer
planet and a stalled inner planet with a constant semi-major axis
(see the first 5000 yrs in the top panel of 
Fig.~\ref{fig:aeo-tw6y}).

During the inward migration process the eccentricity of the outer
planet remains small. As can be seen from Fig.~\ref{fig:sigma} there
is always a sufficient amount of matter in the immediate vicinity 
(co-orbital region) of the planet to ensure damping of the eccentricity. 
For a 5 Jupiter mass planet at $r=6.63$AU the Roche-lobe size is $0.78$ AU,
which is indicated by the radius of the drawn circle. The arrows denote
the position of the inner and outer 2:1 Lindblad resonances.
\begin{figure*}
\begin{minipage}{0.48\linewidth}
\resizebox{\hsize}{!}{%
\includegraphics{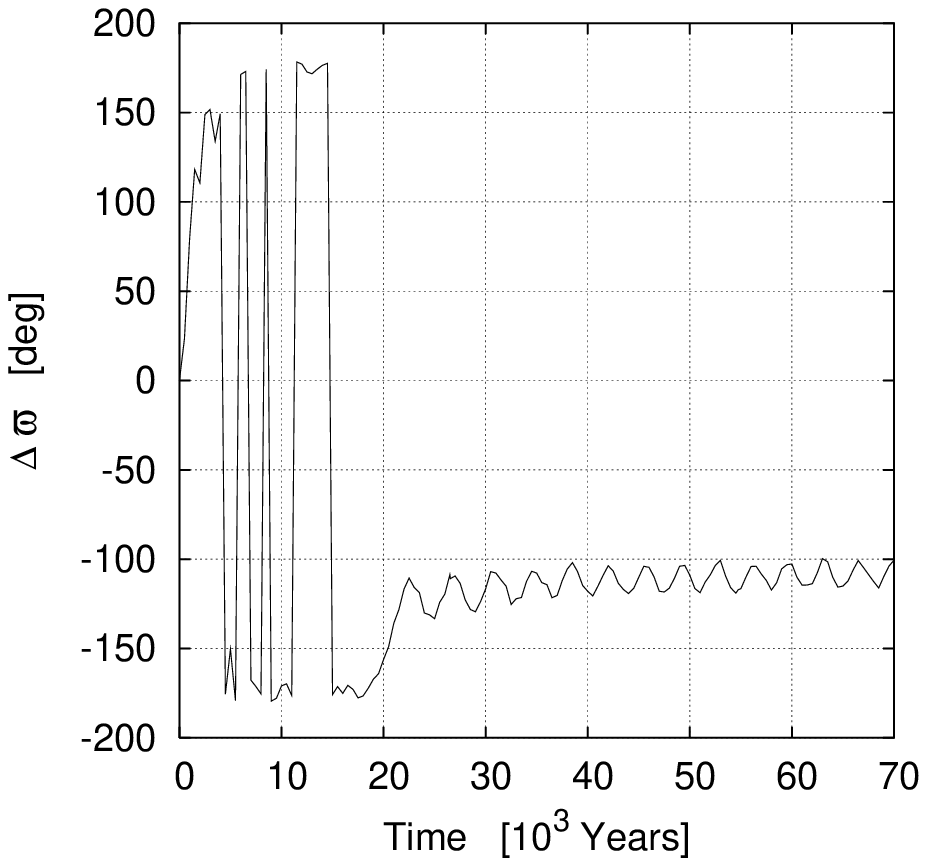}}
\end{minipage}
 \hfill \begin{minipage}{0.48\linewidth}
\resizebox{\hsize}{!}{%
\includegraphics{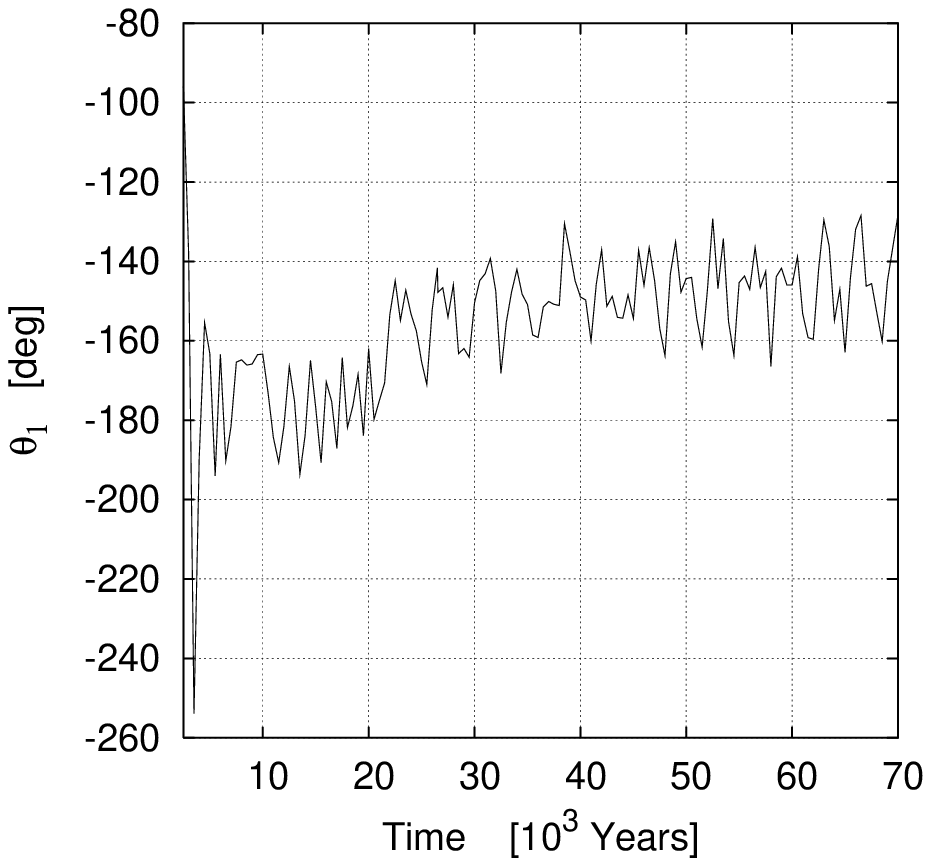}}
\end{minipage}
\caption{
Evolution of $\Delta \pomega$ and $\Theta_1$
in model C.
{\bf Left:} The difference in the direction of the periastrae,
$\Delta \pomega = \pomega_2 - \pomega_1$ (in degrees) for
both planets versus time (in thousand years).
{\bf Right:} The resonant angle for the 3:1 resonance
$\Theta_1 = 3 \lambda_2 - \lambda_1 - 2 \pomega_1$ (in degrees)
as a function of time. 
}
\label{fig:o2-l31.tw6y}
\end{figure*}

The decrease in separation between the planets increases their gravitational
interactions.
Once the ratio of the planets' orbital periods 
has reached a ratio of two integers, i.e. they are close to a
mean motion resonance, 
resonant capture of the inner planet by the outer one may ensue.
Whether or not this does actually happen depends on the physical
conditions in the
disk (e.g. viscosity) and the orbital parameters of the planets.
If the migration speed is too large, for example, there may not be enough
time to excite the resonance, and the outer planet will continue 
migrating inward \citep[e.g.][]{1999MNRAS.304..185H}.
Also, if the initial eccentricities are too small, then
there may be no capture, 
particularly for second-order resonances such as the 3:1 resonance
\citep[see e.g.][]{1999ssd..book.....M}.
For more details on capture probability see
section (\ref{subsubsec:capture}) below.
\subsubsection{3:1 Resonance: Model C}
\label{subsec:Model-C}
The typical time evolution of the semi-major axis ($a$),
eccentricity ($e$) and direction of the periastron ($\pomega$)
are displayed in Fig.~\ref{fig:aeo-tw6y} for the standard model C.
The planets were initialized with zero eccentricities at 
distances of 4 and 10 AU in a disk with partially cleared gaps.

In the beginning, after the inner gap has completely cleared,
only the outer planet migrates inward, and the eccentricities
of both planets remain relatively small, 
($\lapp 0.02$).
After about 5000 yrs the outer planet has reached a semi-major axis
with an orbital period three times that of the inner planet.
The periodic gravitational forcing leads to
the capture of the inner planet into a 3:1 resonance with the outer one.
This is indicated by the dotted reference line (labeled 3:1)
in the top panel of Fig.~(\ref{fig:aeo-tw6y}), which
marks the location of the 3:1 resonance with respect to the inner planet.

We summarize the following important features of the evolution
after resonant capture:
\begin{enumerate}
\item[a)]
In the course of the subsequent evolution, the outer planet,
which is still driven inward by the
outer disk material, forces the inner planet to also
migrate inwards.
Both planets migrate inward simultaneously, always retaining their
resonant configuration.
Consequently, the migration speed of the outer planet slows down,
and their radial separation declines.
\item[b)]
Upon resonant capture the eccentricities of both planets grow
initially very fast before settling into an oscillatory
quasi-static state
which changes slowly on a secular time scale.
This slow increase of the eccentricities on the longer time scale
is caused by the growing gravitational forces between the planets,
due to the decreasing radial distance of the two planets
on their inward migration process.
\item[c)]
The ellipses/periastrae of the planets rotate at a constant, retrograde 
angular speed $\dot{\pomega}$. Coupled together by the
resonance, the apsidal precession rate $\dot{\pomega}$ for both planets is identical,
which can be inferred from the parallel lines
in the bottom panel of Fig.~\ref{fig:aeo-tw6y}. The orientation
of the orbits is phase-locked with a constant separation
$\Delta \pomega = \pomega_2 - \pomega_1$.
The rotation period of the ellipses (apsidal lines) is slightly longer
than the oscillation period of the eccentricities.
\end{enumerate}
\begin{figure}[ht]
\begin{center}
\resizebox{0.98\linewidth}{!}{%
\includegraphics{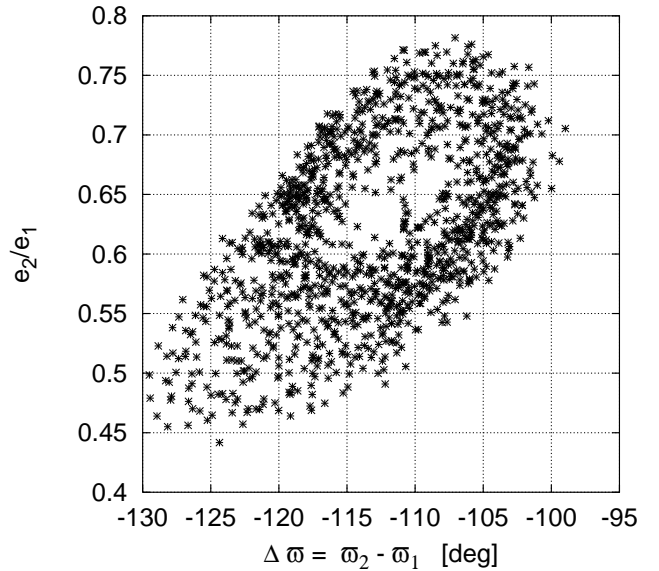}}
\end{center}
\caption{
Eccentricity ratio $e_2/e_1$ of the outer and inner
planet, versus periastron difference $\Delta \pomega$
for model C. 
The data points are spaced equally in time with a distance of approximately
$\delta t = 23$ yrs.
Shown is a section of the evolution
of model C, from 26,000 to 55,000 yrs,
which covers about 7 libration periods.
\label{fig:c1-tw6y}
}
\end{figure}
\begin{figure}[ht]
\begin{center}
\resizebox{0.98\linewidth}{!}{%
\includegraphics{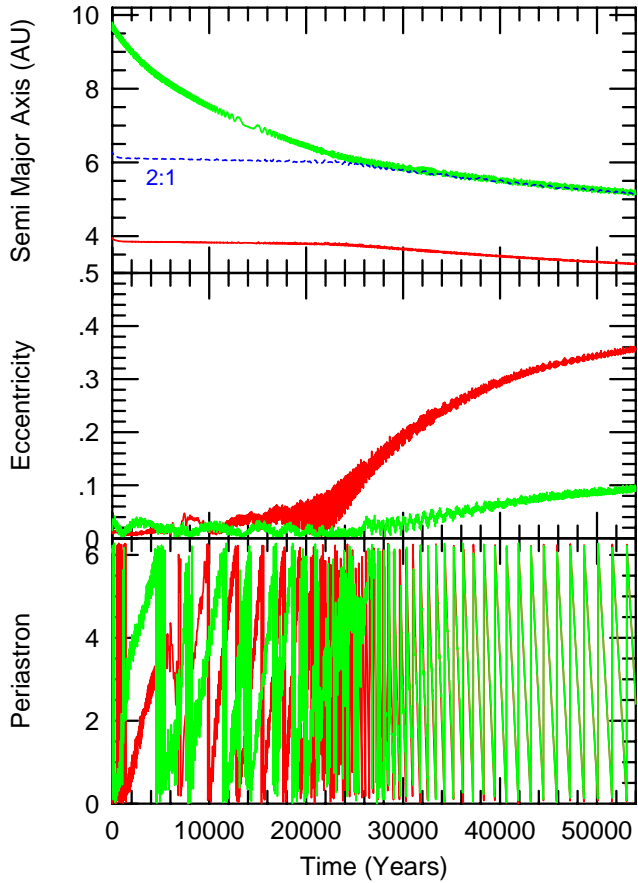}}
\end{center}
  \caption{The semi-major axis, eccentricity and position angle of the
orbital periastron versus time for Model D.
The only parameter different from the first model C is the
lower viscosity.
Here, the outer planet passes through the 3:1 resonance and 
captures the inner planet finally into a 2:1 configuration.
The dotted reference line (labeled 2:1)
marks the location of the 2:1 resonance with respect to the inner planet.
Upon resonant capture, the eccentricities grow
and the two orbits librate retrograde
with a fixed relative orientation of $\Delta \pomega = 0^o$.
}
\label{fig:aeo-tw6z}
\end{figure}

The capture into resonance and
the subsequent libration of the orbits in model C
is illustrated further in Fig.~\ref{fig:o2-l31.tw6y}.
As suggested in Fig.~\ref{fig:aeo-tw6y} (bottom panel)
the periastrae begin to align upon capture in the 3:1 resonance.
Initially, during the phase when the eccentricities are still rising
(between 5 and 20 thousand yrs), the
difference of the periastrae settles intermediately 
to $\Delta \pomega \approx 180^o$ (see
Fig.~\ref{fig:o2-l31.tw6y}, left panel). Then, upon saturation
after about 20,000 yrs, the system re-adjusts and
eventually establishes itself at
$\Delta \pomega \approx 107^o$, with a libration amplitude of about
$7^o$. 
The right panel of Fig.~\ref{fig:o2-l31.tw6y} shows the time evolution
of the resonant angle
$\Theta_1 = 3 \lambda_2 - \lambda_1 - 2 \pomega_1$.
Here $\lambda_1$ and $\lambda_2$ denote the mean longitudes
of the inner and outer planet, respectively.
Initially, $\Theta_1$ settles to $180^o$ as well,
and re-adjusts after 20,000 yrs to $-145^o$ with
a libration amplitude of $\pm 15^o$.
This behavior of an initially symmetric alignment of $\Delta \pomega$
{\it and} $\Theta_1$ at about $180^o$ followed by a later change
to $\Delta \pomega \approx \pm 110^o$ and $\Theta_1 \approx \pm 145^o$ is
characteristic for {\it all} our models which show captures into 3:1 resonance,
independent of the physical parameters (see Tab.~\ref{tab:results}).

This behavior can be understood by an analysis of the interaction
Hamiltonian for resonant systems \citep{2002astro-ph..0210577}.
By minimizing the interaction energy, the equilibrium values for
$\Delta \pomega$ and $\Theta_1$ can be obtained as a function of the
mass and eccentricity of the two planets. As shown by
\citet{2002astro-ph..0210577}, when the eccentricity of the inner
planet is small ($e_1 \lapp .12$) the equilibrium values
of both resonant angles are exactly $180^o$. For higher eccentricity
though, the equilibrium values of $\Delta \pomega$ and $\Theta_1$ shift  
to $115^o$ and $210^o$. In our numerical simulations we find exactly
this behavior. Initially, upon entering the resonant configuration the
eccentricities are small and the two angles both adjust to $180^o$.
Later they readjust to new values as the eccentricities rise
above the critical value.

In the subsequent longterm evolution after 20,000 yrs,
the system settles into a quasi-equilibrium situation where the
eccentricities oscillate with a period of about
3750 yrs. In the $e_2/e_1$ versus $\Delta \pomega$ diagram
(Fig.~\ref{fig:c1-tw6y}), this
phenomenon is demonstrated by the circular distribution of data points
around the equilibrium values.

\subsubsection{2:1 Resonance: Model D}
\label{subsec:Model-D}
\begin{figure*}[ht]
\begin{minipage}{0.48\linewidth}
\resizebox{\hsize}{!}{%
\includegraphics{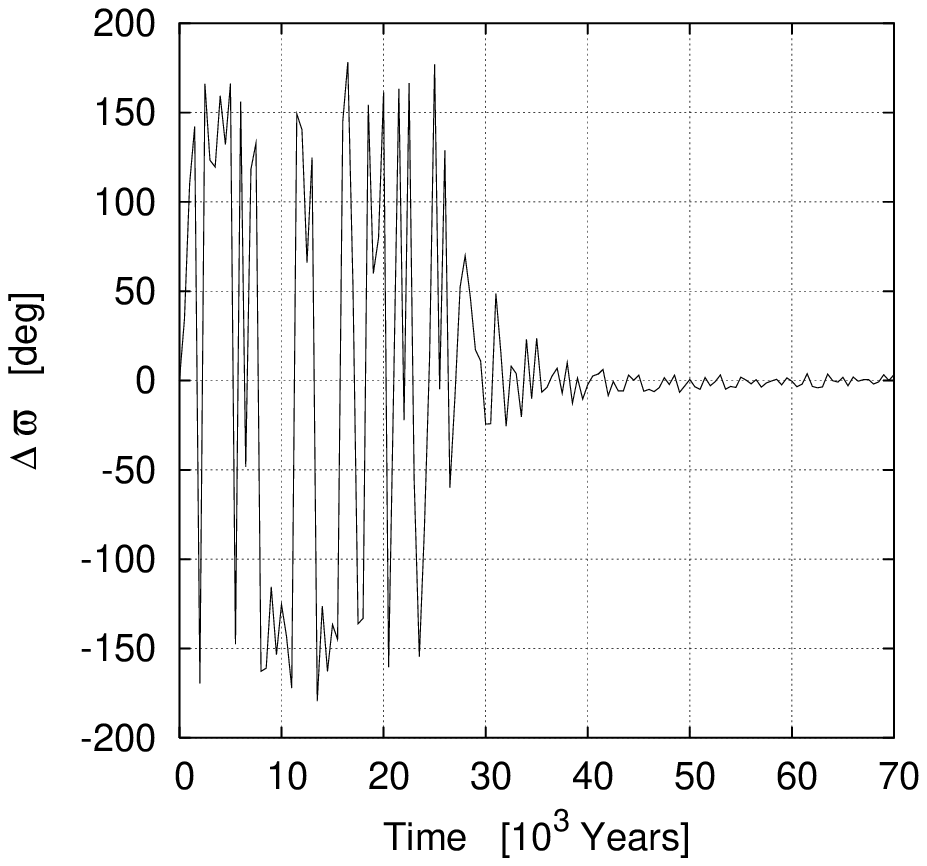}}
\end{minipage}
 \hfill \begin{minipage}{0.48\linewidth}
\resizebox{\hsize}{!}{%
\includegraphics{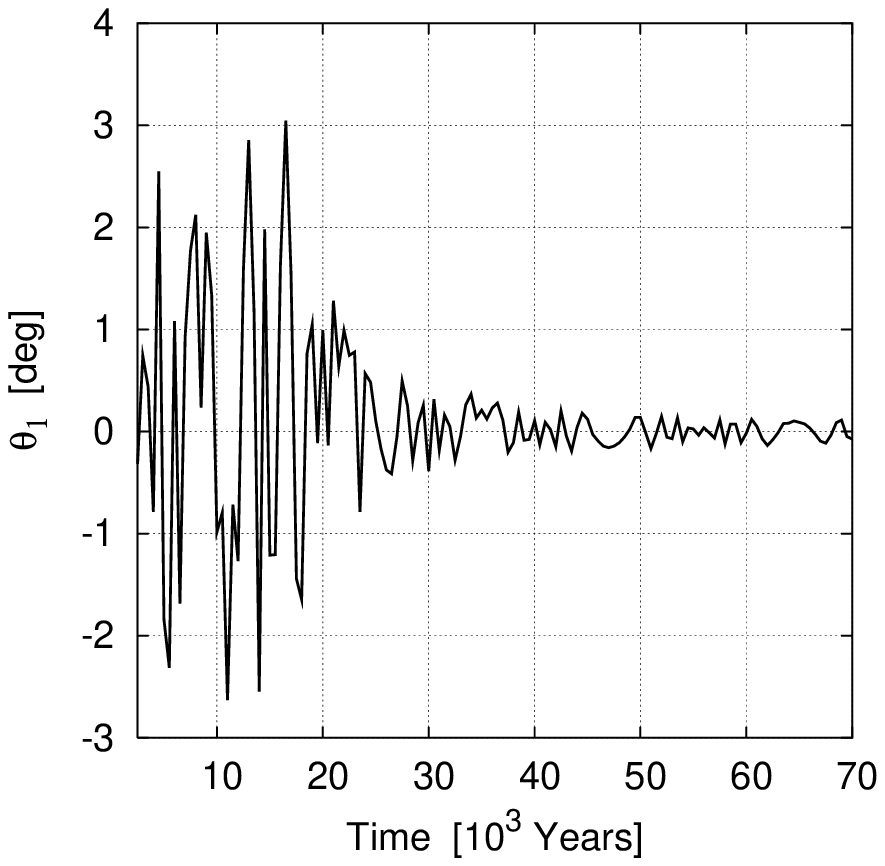}}
\end{minipage}
\caption{
Results for model D.
{\bf Left:} The difference in the direction of the periastrae,
$\Delta \pomega = \pomega_2 - \pomega_1$ (in degrees) for
the two planets versus time (in thousand years).
{\bf Right:} The resonant angle for the 2:1 resonance 
$\Theta_1 = 2 \lambda_2 - \lambda_1 - \pomega_1$ (in degrees),
as a function of time. 
}
\label{fig:o2-l31.tw6z}
\end{figure*}
\begin{figure}[ht]
\begin{center}
\resizebox{0.98\linewidth}{!}{%
\includegraphics{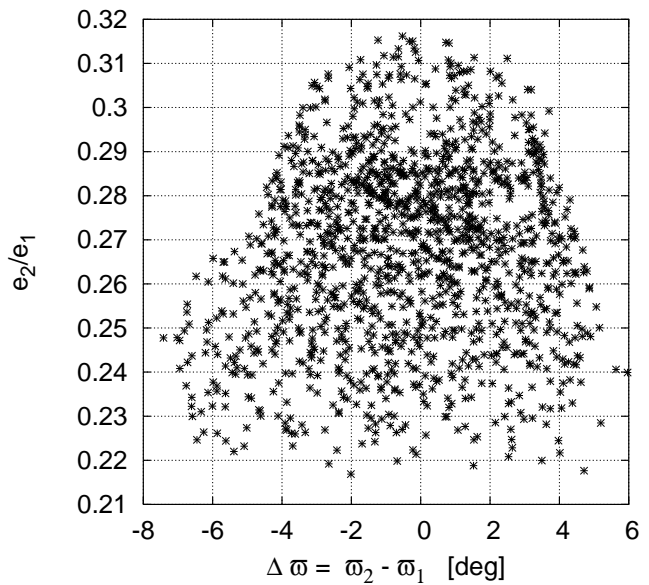}}
\end{center}
\caption{
Eccentricity ratio $e_2/e_1$ of the outer and inner
planet, versus periastron difference $\Delta \pomega$
for model D.
The data points are spaced equally in time with a distance of approximately
$\delta t = 22$ yrs.
Shown is a section of the evolution
of model D, from 43,000 to about 74,000 yrs.
\label{fig:c1-tw6z}
}
\end{figure}

In comparison to the previous model C the only difference in this model D
is the value of the viscosity coefficient.
Three times less viscosity ($\alpha=3.3 \cdot 10^{-3}$)
results in a little bit slower migration speed. 
For this model the evolution did not end up in the 3:1, but rather the
2:1, resonance. In Figure~\ref{fig:aeo-tw6z}, the evolution of
$a$, $e$ and $\pomega$ is displayed. The eccentricities
show a small 'kink' as the outer planet reaches the 3:1
resonance, but migration continues past the resonant location.
Later, at about 26,000 yrs, capture occurs in the 2:1 resonance,
leading to perfectly aligned orbits. Both resonant angles,
$\Delta \pomega$ and $\Theta_1$, are nearly zero with a very small
libration amplitude. 
The $e_2/e_1$ versus $\Delta \pomega$ plot 
(Fig.~\ref{fig:c1-tw6z}) shows
the small variations in the eccentricities and the small
libration amplitude (cf. Fig.~\ref{fig:c1-tw6y}).
Using the same model parameters, we ran a nearly identical simulation
in which the initial density was disturbed randomly by 1\%. 
With just this small change in the initial conditions,
capture into 3:1 resonance was successful 
(see also Tab.~\ref{tab:results} and Sect.~\ref{subsubsec:capture}).
\subsection{Damped N-Body Calculations}
\label{subsec:Nbody}
As a test of the damped N-body model described above
(Sect.~\ref{subsec:model-Nbody}),
we calculate the evolution of all models C-G
using the 3-body method
and compare the results to the full hydrodynamic evolutions.
As outlined above, 
the prescribed damping formula (Eq.~\ref{eq:adot}) is used
to directly alter the semi-major axis $a$ and eccentricity $e$ of 
the outer planet only.
\begin{table}[ht]
\caption{Fit parameters of the N-body computations, obtained
through comparison with the full hydro simulations for models
C-G. Listed are the model name, the initial damping time scale $\tau_0$,
the slow down of the damping $\beta$ (see Eq.~\ref{eq:adot}),
and the eccentricity damping $K$ (see Eq.~\ref{eq:edot}).
}
\label{tab:nbody}
\begin{center}
\begin{tabular}{cccc}
 \hline
Model &   $\tau_0$ [yrs]  &   $\beta$   &   $K$  \\ 
 \hline
  C    &   17,500  &   3.2   &  1.5   \\
  D    &   27,000  &   2.5   &  1.5   \\
  E    &   38,000  &   2.5   &  1.5   \\
  F    &   19,500  &   1.5   &  1.5   \\
  G    &   33,500  &   1.0   &  2.5   \\
 \hline
\end{tabular}
\end{center}
\end{table}
\begin{figure}[ht]
\begin{center}
\resizebox{0.98\linewidth}{!}{%
\includegraphics{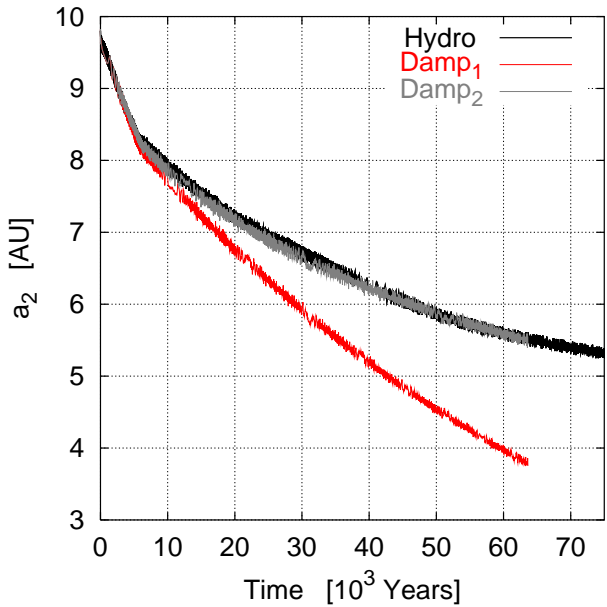}}
\end{center}
\caption{
The evolution of the semi-major axis of the outer planet $a_2$,
comparing a hydrodynamic model with two N-body calculations.
The black curve (labeled Hydro)
is the result of the full hydro model G. The lower dark gray curve
is the damped N-body model using a constant damping $\tau=\tau_0$ in
Eq.~\ref{eq:adot}. The upper lighter gray curve (on top of the black
one) is for time-dependent damping with a non-zero value
of $\beta = 1.0$, resulting in a much better match with the hydro model.
\label{fig:xx7p}
}
\end{figure}

\begin{figure}[ht]
\begin{center}
\resizebox{0.98\linewidth}{!}{%
\includegraphics{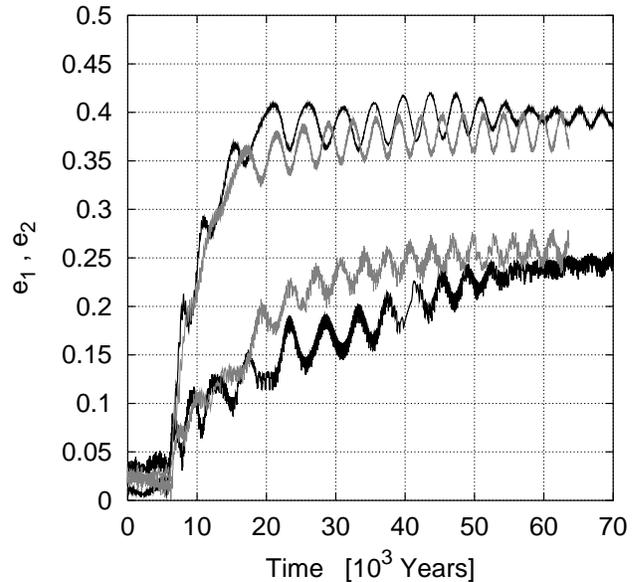}}
\end{center}
\caption{
The evolution of the eccentricities of the inner planet (upper curves)
and the outer planet (lower curves) for the full hydro model G
(black curves covering the whole time range) and the damped N-body model
(light gray), using a damping constant of $K=2.5$.
\label{fig:e7yp}
}
\end{figure}

All the results of the damped N-body models are summarized
in Tab.~\ref{tab:nbody}. The values of the damping constant $K$ cannot
be determined too precisely, as there is some dependence of the
magnitude on the initial eccentricities. Those were chosen to
be about $0.01$ to $0.02$ for all models.

As an example, we display the evolution of model G, in which
$H/r = 0.05$.
To test the damping, we first use 
Eq.~(\ref{eq:adot}) with a constant damping $\tau = \tau_0$, i.e. $\beta=0$.
This refers to the {\it lower} (dark gray) curve in Fig.~(\ref{fig:xx7p}), where
we chose $\tau_0  = 33,500$ yrs.
As can be seen in the figure, a constant damping, even when it
has the correct initial slope,
in the longterm yields too fast a migration of the outer planet.
A time-dependent damping with
$\beta = 1.0$ (light gray curve), 
leads to a much better fit with the hydrodynamic results.
Additionally, we tested our method on published results
for migrating single planets \citep{2000MNRAS.318...18N},
and find good agreement for suitable $\tau_0$ and $\beta$.
Pure exponential fits for $a_2(t)$ with $\beta = 0$
generally do not give satisfactory results.

This decrease in the damping rate as a function of time is a result of the
reduction of mass in the disk, mainly due to the accretion of matter
onto the outer planet.
The mass flow across the gap is small, and the
accretion rate onto the inner planet is substantially lower
\citep[see][]{2000MNRAS.313L..47K}. However, if no gas is allowed to accrete
onto the outer planet, a more substantial amount of
gas may flow occur across the gap, leading to a smaller migration rate,
or even outward migration 
\citep{2001MNRAS.320L..55M, 2002A&A...387..605M,
2003astro-ph..0301171}, because
the angular momentum lost by the gap crossing material
will be gained by the planet. 

Besides $a$, we also compared the evolution of the eccentricity $e$
between the two approaches.
In Fig.~(\ref{fig:e7yp}) we display the eccentricity
evolution of the full hydro and the damped 3-body case for model G.
Despite some differences which we attribute
to the unknown eccentricity damping mechanism, the overall
agreement is reasonable. 
For a given semi-major axis damping rate,
the final values obtained for $e_1$ and $e_2$ at larger times depend
on the initial values for the eccentricities
and the amount of eccentricity damping. In this case we used an initial
$e(t_0) = 0.01$ for both planets and an eccentricity damping factor
$K=2.5$, i.e. a slightly shorter
damping time scale for eccentricity as for semi-major axis.
For all models we find that the eccentricity damping rate is of
the same order as the semi-major axis damping, i.e. $K = {\cal{O}}(1)$.
This finding is in contrast to \citet{2002ApJ...567..596L}
who determined a much shorter eccentricity damping time, based
on models for GJ~876. 

There are several possible reasons for this
difference. Since the eccentricity damping of a planet is caused by
material in the co-orbital region close to the planet, the treatment
of this region in the models may have some effect on the results. 
In particular, the mass accretion onto the planet, the smoothing of
the gravitational potential, and the numerical resolution may each
play a significant role here. 
However, the simulations by 
\citet{2002ApJ...567..596L} are not based on clear physical model of the
damping, but rather use an ad hoc prescription. Fitting to the observed
case of GJ~876 yields a high value of $K$. On the other hand, it will be
difficult to model the system HD~82943 with $K=100$ because of the high
observed eccentricities.
\begin{figure}
\begin{center}
\resizebox{0.98\linewidth}{!}{%
\includegraphics{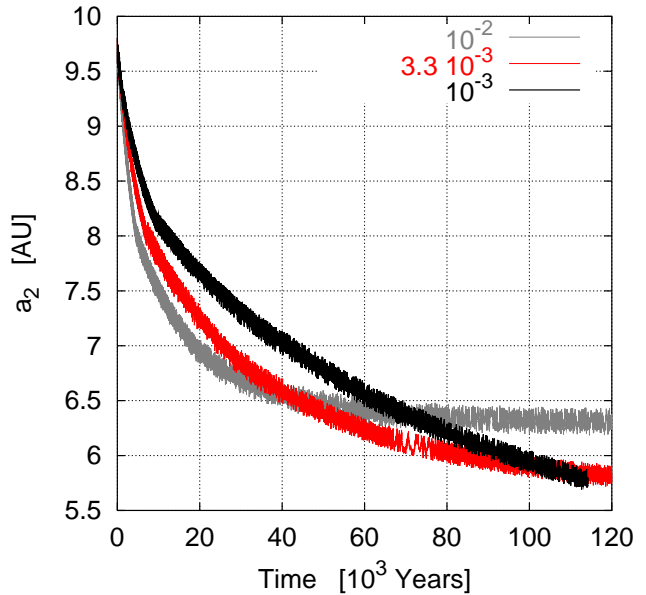}}
\end{center}
  \caption{The time evolution of the outer planet's  semi-major axis 
   for three
   models (C-E) with different viscosities. The corresponding $\alpha$ values 
   are given in the legend
   \label{fig:a6visc}
    }
\end{figure}
\subsection{Dependence on the Physical Parameters}
\label{subsec:analysis}
After describing the major effects of resonant capture and evolution
we focus now on the dependence on the physical parameters.
An overview of the results for all models is given in Tab.~\ref{tab:results}.
\begin{table}[ht]
\caption{Results of the full hydrodynamical
computations for the longterm evolution.
Listed are the model name, the resonance in which the system is 
captured, the eccentricity ratio $e_2/e_1$, the separation of the
periastrae $\Delta \pomega$, the resonant angle $ \Theta_1 $,
and the speed of the apsidal precession $\dot{\pomega}$. The models
labeled with (V) are still evolving with time. For model D two
cases (caught in different resonances) are presented.
}
\label{tab:results}
\begin{tabular}{cclrrl}
 \hline
Model &  Res.  &  $e_2/e_1$ &  $\Delta \pomega$  & $ \Theta_1 $
     & $\dot{\pomega}$  \\
     &            &    &   [deg]     &  [deg]  &  [rad/yr]    \\
 \hline
  A    &   3:1   &  1.0   &  -110   &  -140  &  - 0.0015   \\
  B    &   3:1   &  0.7   &  -120   &  -150  &  - 0.0015   \\
 \hline
  C    &   3:1   &  0.73  &  -107  & -145  &  - 0.0014   \\
  D1   &   2:1   &  0.35  &   0    &   0   &  - 0.0033    \\
  D2   &   3:1   &  0.70  &  +110  &  147  &  - 0.0015 (V) \\
  E    &   3:1   &  0.73  &  -107  & -140  &  - 0.0010     \\
  F    &   5:2   &  1.1   &  +180  &   0   &  - 0.0033     \\
  G    &   3:1   &  0.65  &  +110  & +150  &  - 0.0022 (V) \\
 \hline
  X    &   2:1   &  0.24  &    0   & 0     & - 0.0021    \\
 \hline
\end{tabular}
\end{table}
\subsubsection{Viscosity}
\label{subsubsec:viscosity}
As the disk viscosity, parameterized here through the 
standard $\alpha$-parameter,
determines the overall evolution of the disk,
it is to be expected that its magnitude influences the
longterm evolution of the planetary system as well.
Starting from the standard model C ($\alpha =10^{-2}$)
we performed additional runs using
different values for the viscosity parameter $\alpha$;
model D has $\alpha =3.3 \cdot 10^{-3}$
and model E has $\alpha = 10^{-3}$.
The semi-major axis evolution for these models is shown in
Fig.~\ref{fig:a6visc}. For a single planet, the migration rate depends on the
value of the viscosity \citep{2000MNRAS.318...18N}. Hence, the
initial migration rates (during the first 10,000 yrs) are reduced for
smaller $\nu$.  The initial migration speeds for each model, as seen
in Fig.~\ref{fig:a6visc}, are quantified as $\tau_0$ in
Tab.~\ref{tab:nbody}. 

Upon capture of the inner planet, the migration rate of the outer slows down.
Contrary to the expectation that smaller viscosities yield slower
migration, we find that for these resonantly driven double-planet systems,
the speed of migration is eventually faster for smaller
viscosity coefficients.
When the viscosity is lower, the migration slows down less rapidly
(this would correspond to a lower $\beta$-value in the N-body method).
This time-dependence of the damping is linked to the assumed
mass accretion onto the planet. The accretion, which is larger for
higher viscosity, lowers the mass of the
disk with time and reduces the effective torques, such that 
migration drops off more rapidly for higher viscosity.
The ratio of semi-major axis damping to eccentricity damping
($K$), however, does not appear to depend on the value of the viscosity.
In all three cases we find capture into
the 3:1 resonance (see Tab.~\ref{tab:results}), 
although this result is somewhat indeterminate. 
For the value of $\alpha = 3.3 \cdot 10^{-3}$, we
also find a case in 2:1.
\subsubsection{Temperature}
\label{subsubsec:temperature}
\begin{figure}
\begin{center}
\resizebox{0.98\linewidth}{!}{%
\includegraphics{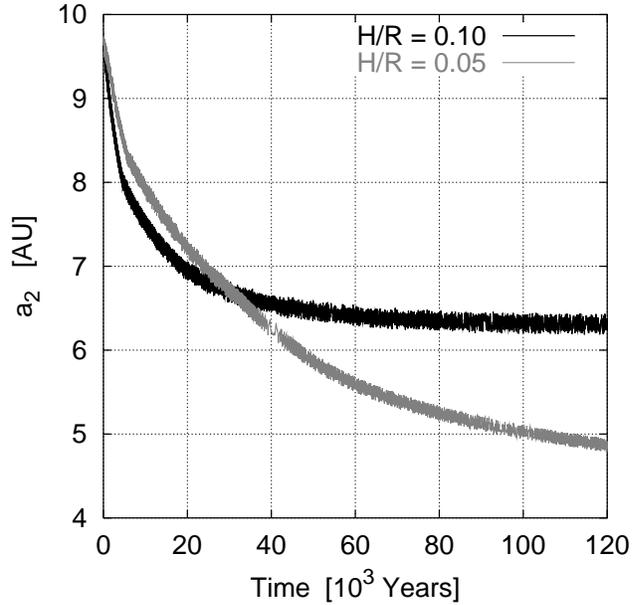}}
\end{center}
  \caption{The evolution of the semi-major axis
   of the outer planet $a_2$ for two
   models (C, G) with different vertical disk scale heights,
   as indicated in the legend.
   \label{fig:a6h}
    }
\end{figure}

Similarly to the viscosity, it may be expected that the scale height
(temperature) of the disk will also influence the migration
process of planets. 
Results for single planets \citep{1999MNRAS.303..696K} have shown
that for larger $H/r$ the gap is less cleared because the
larger pressure gradient `pushes' material into the gap.
The higher density near the gap edges leads to a
faster migration speed. This effect is indeed clearly seen in the early
evolution of the planetary systems C \& G, where the decline in
the semi-major axis of the outer planet $a_2$  is faster for
higher temperatures (see Fig.~\ref{fig:a6h}).

Upon capture however, the evolution behaves similarly to the
case of varying viscosity. Those systems with higher $H/r$ now have the
lowest migration speed.  Similarly to the models with lower viscosity,
smaller $H/r$ leads to lower accretion, higher disk mass, and hence
more material pushing on the planets.
As the migration slows down less for lower $H/r$,
the overall evolution time scale becomes so long that it takes much
more than $10^5$ yrs for 
the system to reach an equilibrium state.
The intermediate model with $H/R=0.075$ is not displayed as is went into
a 5:2 resonance.
\begin{figure}
\begin{center}
\resizebox{0.98\linewidth}{!}{%
\includegraphics{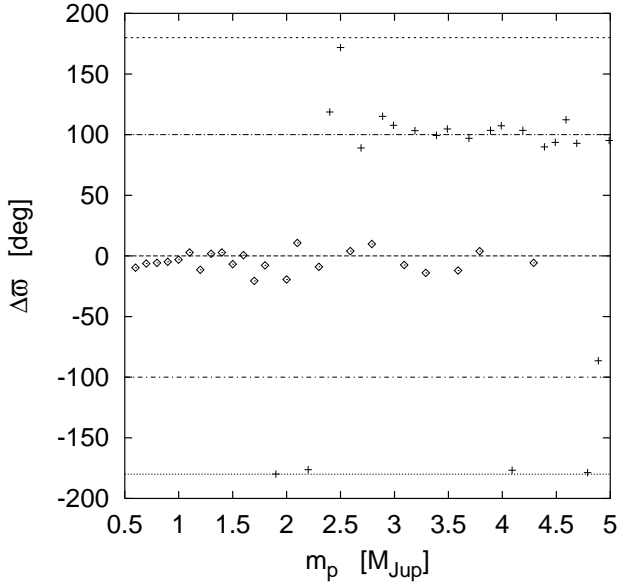}}
\end{center}
  \caption{Results of a sequence of damped N-body simulations. 
   Plotted is the difference of the periastrae
   $\Delta \pomega = \pomega_2 - \pomega_1$ of the two planets after
   capture into resonance versus planet
   mass, where $m_p = m_2 = m_1$. 
   The diamonds indicate capture in 2:1 resonance, while the plus
   signs are for 3:1 resonance.
   The other parameters are fixed, as described in the tex.
   The horizontal
   lines indicate values of $0$, $\pm 100^o$ and $\pm 180^o$.
   \label{fig:mass}
    }
\end{figure}
\subsubsection{Capture Probabilities}
\label{subsubsec:capture}
The main results of the full hydrodynamic calculations,
as summarized in Tab.~\ref{tab:results}, show
that for the same physical setup, capture in different resonances may occur.
As a three body system is intrinsically chaotic, this indeterminate behavior
may be expected. 

Nevertheless, by running a whole sequence of fast damped N-body models,
we can investigate what conditions determine the principle final outcome.
The standard setup
consists of two planets of $1 M_{Jup}$ each, placed initially at 4 and 12 
AU from the central $1 M_\odot$ star. The initial eccentricities
are $0.02$.
As in all previous models, only the orbit of the outer planet is damped,
using in these cases a damping time scale $\tau_0 = 20,000$ yrs.
We fix the damping constants to $\beta = 1.0$ and $K=1.0$. 
Starting from this standard case,
we vary the damping time scale $\tau_0$, the
initial eccentricity $e_2$, and mass of the outer planet $m_2$,
while keeping always identical planet masses, $m_1 = m_2$.

In the standard case, the planets are caught in a 2:1
resonance. Varying the time scale $\tau_0$ from 10,000 to 50,000 yrs,
there is no capture into higher resonances.
Upon variation of eccentricity $e_2$ from 0.0 to 0.5, we find that
for $e_2 < 0.25$ capture occurs always into 2:1, while
for larger $e_2$ higher order resonances (primarily 3:1)
are possible, however with no definite outcome.
The influence of the planet mass is illustrated in Fig.~\ref{fig:mass},
where the resonance type (2:1 diamonds, 3:1 plus signs)
is shown as a function of planet mass (with $m_1 = m_2$).
For small planets with $m_p < 1.7 M_{Jup}$, capture occurs robustly
into the 2:1 resonance. 
Higher resonances are possible only for $m_p$ larger than
this value. However, due to the chaotic nature of the problem
the exact outcome for a particular $m_p$ is not predictable.
This is in agreement with the hydrodynamic simulations where
we also find capture in different resonances just by perturbing the
initial density slightly (compare models D1, D2).
\subsubsection{The Resonant Angles}
\label{subsubsec:angles}
In the majority of hydrodynamic models, the planets catch each other in a 3:1
resonance. As demonstrated above, this is mainly a consequence of the
large masses chosen for the planets. For all models the resonant angles settle
to $| \Delta \pomega | \approx 110^o $ and $|\Theta_1 | \approx 145^o$, with
some scatter of about $\pm 10^o$. For the damped N-body models 
we checked the values of the resonant angles as well, finding that all 2:1
resonances settle into the complete symmetric configuration
$\Delta \pomega = \Theta_1 = 0$ (Fig.~\ref{fig:mass}).
For the 3:1 resonances we see anti-symmetric
configurations with anti-aligned periastrae, $|\Delta \pomega | = 180^0$
and $\Theta_1 = 0$ for masses
around $m_p = 2 M_{Jup}$, (when the first 3:1 cases begin to occur),
while for all larger masses we find preferentially
the previous non-symmetric configurations. This behavior is again an
indication of a bifurcation in the stability properties of resonant
systems, as claimed by \citet{2002astro-ph..0210577}.
With additional damped N-body calculations we also find
that systems entering a 5:2 resonance (not shown)
exhibit a typical anti-symmetric behavior $|\Delta \pomega | = 180^0$
and $\Theta_1 = 0^o$.
%
\section{Summary and Conclusion}
\label{sub:summary}
We have performed full hydrodynamical calculations simulating the
joint evolution of a pair of protoplanets together with the surrounding 
protoplanetary disk from which they originally formed.
The focus lies on massive planets in the range of a few Jupiter masses.
For the disk evolution we solve the Navier-Stokes equations, and the
motion of the planets is followed using a 4th order Runge-Kutta scheme,
which includes their mutual interactions as well as the star and disk's
gravitational fields.
These results were compared to simplified (damped) N-body computations,
where the gravitational influence of the disk is modeled through analytic
damping terms applied to the semi-major axis and eccentricity.
 
We find that both methods yield comparable results, if the damping
constants in the simplified models are adjusted properly.
The mass reduction of the disk with time, due for example
to mass accretion onto the planet, or possible mass flow across the
outer planet's gap can be modeled satisfactorily through a
damping time scale, which depends linearly on time.
The eccentricity damping was always chosen to be a constant multiple $K$
of the semi-major axis damping. In this case we find that $K$ must be
of order unity to match the hydrodynamic models. 

However, fitting N-body models to the observed parameters of GJ~876
requires a high $e$-damping with typically
$K = 100$ \citep[see also][]{2002ApJ...567..596L},
relatively independent of the functional
behavior of $a_2(t)$. Reasons for this discrepancy may lie in the
simplified hydrodynamical model, which uses a fixed equation of state,
a simple treatment of the planetary structure, and only an approximate
model of the torques acting on the planet. 
Also, eccentricity damping is dominated by material
close to the planet; the insufficient numerical grid resolution near
the planet may smear out the damping forces.
In addition, the accretion process of matter onto the planet
is reducing the mass in the
co-orbital region which lowers the eccentricity damping.
The simplified assumption of a constant value of $K$ needs to
be checked.
More detailed hydrodynamical
models may help to resolve this discrepancy in the future.
An alternative explanation for the low eccentricities in GJ~876 compared
to our hydrodynamic simulations is that further evolution of the
eccentricities occurs in the system after planet and gap formation.
The planet eccentricities may be further modified as the disk
dissipates and its resulting eccentricity forcing gradually declines.
On the other hand, the assumption of a constant value of $K$ in the
N-body models in the computations by \citet{2002ApJ...567..596L}
is also not based on any detailed hydrodynamic model but rather
assumed ab initio. In more general models, this
will have to be relaxed.

The case HD~82943 is also not easy to model as the eccentricities for
both planets are very large, which turned out to be very difficult
to capture with N-body models, even with very low damping. 
The problem here lies in the stability of the resulting system. 
All test computations with constant values of $K$ eventually
led to unstable systems.
Compared to GJ~876, the eccentricity damping for HD~82943 
must be orders of magnitude less, if otherwise similar physical
parameters are used.
In order to explain the high eccentricities, 
the inclusion of an additional companion may be necessary.

Despite of the difficulty of the models to obtain the 
observed eccentricities, there are nevertheless
several features of the observed 2:1 planets which are captured
correctly by our simulations: {\it i}) The larger mass
of the outer planet, {\it ii}) the higher eccentricity of the
inner planet, and {\it iii}) the periastrae separation of
$\Delta \pomega = 0^o$. 
These are robust predictions of the hydrodynamic models.

For 3:1 resonances, anti-symmetric 
($\Delta \pomega = 180^o$) and non-symmetric final configurations
are obtained. In the non-symmetric case we found over a range of
models a value of $| \Delta \pomega \approx 110^o|$, which is 
supported by stability analysis \citep{2002astro-ph..0210577}.
In 55~Cnc, the only observed 3:1 case, there are other planets present
in the system, which makes an interpretation using just this simple
treatment questionable.

\begin{acknowledgements}
We would like to thank Doug Lin, Richard Nelson, and John Papaloizou
for many stimulating discussions during the course of this project.
\end{acknowledgements}

\bibliographystyle{aa}
\bibliography{kley1}

\begin{thebibliography}{37}
\expandafter\ifx\csname natexlab\endcsname\relax\def\natexlab#1{#1}\fi

\bibitem[{{Bate} {et~al.}(2003){Bate}, {Lubow}, {Ogilvie}, \&
  {Miller}}]{2003astro-ph..0301154}
{Bate}, M.~R., {Lubow}, S.~H., {Ogilvie}, G.~I., \& {Miller}, K.~A. 2003, {\tt
  astro-ph/0301154}, \mnras, in press

\bibitem[{{Beauge} {et~al.}(2002){Beauge}, {Ferraz-Mello}, \&
  {Michtchenko}}]{2002astro-ph..0210577}
{Beauge}, C., {Ferraz-Mello}, S., \& {Michtchenko}, T.~A. 2002,
  astro-ph/0210577, \apj, submitted

\bibitem[{{Bois} {et~al.}(2003){Bois}, {Kiseleva-Eggleton}, {Rambaux}, \&
  {Pilat-Lohinger}}]{2003astro-ph..0301528}
{Bois}, E., {Kiseleva-Eggleton}, L., {Rambaux}, N., \& {Pilat-Lohinger}, E.
  2003, {\tt astro-ph/0301528}, \apj, submitted

\bibitem[{{Bryden} {et~al.}(1999){Bryden}, {Chen}, {Lin}, {Nelson}, \&
  {Papaloizou}}]{1999ApJ...514..344B}
{Bryden}, G., {Chen}, X., {Lin}, D.~N.~C., {Nelson}, R.~P., \& {Papaloizou},
  J.~C.~B. 1999, \apj, 514, 344

\bibitem[{{Bryden} {et~al.}(2000){Bryden}, {R{\' o}{\. z}yczka}, {Lin}, \&
  {Bodenheimer}}]{2000ApJ...540.1091B}
{Bryden}, G., {R{\' o}{\. z}yczka}, M., {Lin}, D.~N.~C., \& {Bodenheimer}, P.
  2000, \apj, 540, 1091

\bibitem[{{Butler} {et~al.}(1999){Butler}, {Marcy}, {Fischer}, {Brown},
  {Contos}, {Korzennik}, {Nisenson}, \& {Noyes}}]{1999ApJ...526..916B}
{Butler}, R.~P., {Marcy}, G.~W., {Fischer}, D.~A., {et~al.} 1999, \apj, 526,
  916

\bibitem[{{D'Angelo} {et~al.}(2002){D'Angelo}, {Henning}, \&
  {Kley}}]{2002A&A...385..647D}
{D'Angelo}, G., {Henning}, T., \& {Kley}, W. 2002, \aap, 385, 647

\bibitem[{{D'Angelo} {et~al.}(2003){D'Angelo}, {Kley}, \&
  {Henning}}]{2003ApJ...586..540D}
{D'Angelo}, G., {Kley}, W., \& {Henning}, T. 2003, \apj, 586, 540

\bibitem[{{Go{\' z}dziewski} \& {Maciejewski}(2001)}]{2001ApJ...563L..81G}
{Go{\' z}dziewski}, K. \& {Maciejewski}, A.~J. 2001, \apjl, 563, L81

\bibitem[{{Goldreich} \& {Tremaine}(1980)}]{1980ApJ...241..425G}
{Goldreich}, P. \& {Tremaine}, S. 1980, \apj, 241, 425

\bibitem[{{Haghighipour}(1999)}]{1999MNRAS.304..185H}
{Haghighipour}, N. 1999, \mnras, 304, 185

\bibitem[{{Ji} {et~al.}(2003{\natexlab{a}}){Ji}, {Kinoshita}, {Liu}, {Guangyu},
  \& {Nakai}}]{2002astro-ph..0301353}
{Ji}, J., {Kinoshita}, H., {Liu}, L., {Guangyu}, L., \& {Nakai}, H.
  2003{\natexlab{a}}, in Proceedings of IAU 189 Colloquium, Sept. 2002,
  Nanjing, P.R. China, in press, astro--ph/0301353

\bibitem[{{Ji} {et~al.}(2003{\natexlab{b}}){Ji}, {Kinoshita}, {Liu}, \&
  {Li}}]{2003ApJ...585L.139J}
{Ji}, J., {Kinoshita}, H., {Liu}, L., \& {Li}, G. 2003{\natexlab{b}}, \apjl,
  585, L139

\bibitem[{{Ji} {et~al.}(2002){Ji}, {Li}, \& {Liu}}]{2002ApJ...572.1041J}
{Ji}, J., {Li}, G., \& {Liu}, L. 2002, \apj, 572, 1041

\bibitem[{{Kley}(1998)}]{1998A&A...338L..37K}
{Kley}, W. 1998, \aap, 338, L37

\bibitem[{{Kley}(1999)}]{1999MNRAS.303..696K}
---. 1999, \mnras, 303, 696

\bibitem[{{Kley}(2000)}]{2000MNRAS.313L..47K}
---. 2000, \mnras, 313, L47

\bibitem[{{Kley}(2003)}]{2002astro-ph..0302352}
{Kley}, W. 2003, in Procceedings of IAU 189 Colloquium, Sept. 2002, Nanjing,
  P.R. China, in press, astro--ph/0302352

\bibitem[{{Laughlin} \& {Chambers}(2001)}]{2001ApJ...551L.109L}
{Laughlin}, G. \& {Chambers}, J.~E. 2001, \apjl, 551, L109

\bibitem[{{Lee} \& {Peale}(2002{\natexlab{a}})}]{2002ApJ...567..596L}
{Lee}, M.~H. \& {Peale}, S.~J. 2002{\natexlab{a}}, \apj, 567, 596

\bibitem[{{Lee} \& {Peale}(2002{\natexlab{b}})}]{2002astro-ph..0209176}
{Lee}, M.~H. \& {Peale}, S.~J. 2002{\natexlab{b}}, in Scientific Frontiers in
  Research on Extrasolar Planets, ASP Conference Series, in press,
  astro--ph/0209176

\bibitem[{{Lin} \& {Papaloizou}(1980)}]{1980MNRAS.191...37L}
{Lin}, D.~N.~C. \& {Papaloizou}, J. 1980, \mnras, 191, 37

\bibitem[{{Lin} \& {Papaloizou}(1986)}]{1986ApJ...309..846L}
---. 1986, \apj, 309, 846

\bibitem[{{Lin} \& {Papaloizou}(1993)}]{1993prpl.conf..749L}
{Lin}, D.~N.~C. \& {Papaloizou}, J.~C.~B. 1993, in Protostars and Planets III,
  749--835

\bibitem[{{Lubow} {et~al.}(1999){Lubow}, {Seibert}, \&
  {Artymowicz}}]{1999ApJ...526.1001L}
{Lubow}, S.~H., {Seibert}, M., \& {Artymowicz}, P. 1999, \apj, 526, 1001

\bibitem[{{Marcy} {et~al.}(2001){Marcy}, {Butler}, {Fischer}, {Vogt},
  {Lissauer}, \& {Rivera}}]{2001ApJ...556..296M}
{Marcy}, G.~W., {Butler}, R.~P., {Fischer}, D., {et~al.} 2001, \apj, 556, 296

\bibitem[{{Marcy} {et~al.}(2002){Marcy}, {Butler}, {Fischer}, {Laughlin},
  {Vogt}, {Henry}, \& {Pourbaix}}]{2002ApJ...581.1375M}
{Marcy}, G.~W., {Butler}, R.~P., {Fischer}, D.~A., {et~al.} 2002, \apj, 581,
  1375

\bibitem[{{Marcy} {et~al.}(2003){Marcy}, {Fischer}, , {Butler}, \&
  {Vogt}}]{2002marcy-systems}
{Marcy}, G.~W., {Fischer}, D.~A., , {Butler}, R.~P., \& {Vogt}, S.~S. 2003, in
  Space Science Reviews, in press

\bibitem[{{Masset} \& {Snellgrove}(2001)}]{2001MNRAS.320L..55M}
{Masset}, F. \& {Snellgrove}, M. 2001, \mnras, 320, L55+

\bibitem[{{Masset}(2002)}]{2002A&A...387..605M}
{Masset}, F.~S. 2002, \aap, 387, 605

\bibitem[{{Masset} \& {Papaloizou}(2003)}]{2003astro-ph..0301171}
{Masset}, F.~S. \& {Papaloizou}, J.~C.~B. 2003, {\tt astro-ph/0301171}, \apj,
  in press

\bibitem[{{Murray} \& {Dermott}(1999)}]{1999ssd..book.....M}
{Murray}, C.~D. \& {Dermott}, S.~F. 1999, {Solar system dynamics} (Solar system
  dynamics by Murray, C.~D., 1999)

\bibitem[{{Nelson} \& {Papaloizou}(2002)}]{2002MNRAS.333L..26N}
{Nelson}, R.~P. \& {Papaloizou}, J.~C.~B. 2002, \mnras, 333, L26

\bibitem[{{Nelson} {et~al.}(2000){Nelson}, {Papaloizou}, {Masset}, \&
  {Kley}}]{2000MNRAS.318...18N}
{Nelson}, R.~P., {Papaloizou}, J.~C.~B., {Masset}, F.~.~., \& {Kley}, W. 2000,
  \mnras, 318, 18

\bibitem[{{Snellgrove} {et~al.}(2001){Snellgrove}, {Papaloizou}, \&
  {Nelson}}]{2001A&A...374.1092S}
{Snellgrove}, M.~D., {Papaloizou}, J.~C.~B., \& {Nelson}, R.~P. 2001, \aap,
  374, 1092

\bibitem[{{Tanaka} {et~al.}(2002){Tanaka}, {Takeuchi}, \&
  {Ward}}]{2002ApJ...565.1257T}
{Tanaka}, H., {Takeuchi}, T., \& {Ward}, W.~R. 2002, \apj, 565, 1257

\bibitem[{{Ward}(1997)}]{1997Icar..126..261W}
{Ward}, W.~R. 1997, Icarus, 126, 261

\end{thebibliography}
\end{document}